\documentclass[runningheads]{llncs}

\usepackage[T1]{fontenc}
\usepackage{graphicx}
\graphicspath{{figures/}}
\usepackage{booktabs}
\usepackage{amsmath}
\usepackage{multirow}
\usepackage{makecell}
\usepackage{xcolor}
\usepackage{enumitem}
\usepackage{tikz}
\usetikzlibrary{positioning, arrows.meta}
\usepackage[hidelinks]{hyperref}
\setlist{nosep,leftmargin=*}

\makeatletter
\renewcommand\section{\@startsection{section}{1}{\z@}%
                       {-12\p@ \@plus -3\p@ \@minus -3\p@}%
                       {5\p@ \@plus 2\p@ \@minus 2\p@}%
                       {\normalfont\large\bfseries\boldmath
                        \rightskip=\z@ \@plus 8em\pretolerance=10000 }}
\renewcommand\subsection{\@startsection{subsection}{2}{\z@}%
                       {-10\p@ \@plus -3\p@ \@minus -3\p@}%
                       {3\p@ \@plus 2\p@ \@minus 2\p@}%
                       {\normalfont\normalsize\bfseries\boldmath
                        \rightskip=\z@ \@plus 8em\pretolerance=10000 }}
\renewcommand\paragraph{\@startsection{paragraph}{4}{\z@}%
                       {-6\p@ \@plus -2\p@ \@minus -2\p@}%
                       {-0.5em \@plus -0.22em \@minus -0.1em}%
                       {\normalfont\normalsize\itshape}}
\makeatother

\begin{document}

\title{Your Retriever Already Knows:\\Distribution-Shape QPP for RAG Retrieval Sufficiency}
\titlerunning{Distribution-Shape QPP for RAG Retrieval Sufficiency}

\author{Maty\'{a}\v{s} Vesel\'{y}\inst{1}\orcidID{0009-0006-7994-4494} \and
Michal Pr\r{u}\v{s}ek\inst{1,2}\orcidID{0009-0009-0573-3784} \and
Ji\v{r}\'{i} Franc\inst{1}\orcidID{0000-0002-5761-8675}}
\authorrunning{M. Vesel\'{y} et al.}
\institute{Department of Mathematics, FNSPE CTU in Prague \and
Institute of Information Theory and Automation, Czech Academy of Sciences}

\maketitle

\begin{abstract}
Standard Retrieval-Augmented Generation (RAG) pipelines often
provide no reliable inference-time signal of whether retrieval
succeeded; on ambiguous or out-of-scope queries, generation may
then hallucinate. Motivated by a Czech nuclear-regulator deployment
where data sensitivity precludes third-party LLM APIs, we compare
three Query Performance Prediction (QPP) paradigms for retrieval
sufficiency in RAG: score-based features, a content-based
LLM judge, and a hybrid. On the eight ViDoRe vision domains
(14{,}514~queries), our 24 non-lexical features (GeneralQPP;
15~distribution-shape, 5~query-surface, 4~global) reach a
weighted-average AUROC of 0.856 at 2\,ms per query,
ahead of a classic-QPP literature pool (Classic Full,~0.835)
and well above a local Qwen3.5 LLM judge (0.649, $+0.207$ gap;
$\sim$3000$\times$ faster and cheaper per query). Adding the LLM
judgment as one feature (hybrid) matches S1 on ViDoRe (0.863)
but gains a statistically
significant edge on S\'{U}JB (AUROC~0.911 at Hit@5,
adversarial-detection~0.954; 1{,}510~queries, 500 synthetic
adversarial), at LLM latency.
Rankings agree across datasets (Spearman~$\rho\!=\!0.90$).
Under Leave-One-Domain-Out, S1 drops to~0.706; a 13-feature
LODO-stepwise subset (S1-Lean) recovers to~0.719 ($+0.032$
over the literature pool).

\keywords{Query Performance Prediction \and Retrieval-Augmented Generation \and Retrieval Sufficiency \and Safety-Critical RAG \and Vision Document Retrieval}
\end{abstract}

\section{Introduction}
\label{sec:intro}

Retrieval-Augmented Generation (RAG)~\cite{lewis2020rag} grounds a
language model on documents retrieved from a corpus. Our scope is
\emph{pre-generation retrieval-sufficiency prediction} (narrower
than full answer-quality prediction): without ground-truth labels,
predict whether the top-$k$ context likely contains the answer so
the system can decide at inference time whether to answer,
abstain, or escalate before the LLM hallucinates.

In safety-critical deployments the need is acute: our motivating
Czech nuclear-regulator pipeline (S\'{U}JB) forbids third-party
LLM APIs for data-sensitivity reasons, and incorrect answers carry
real consequences, so a confidence estimator must be fast, cheap,
and local. Can the
\emph{distribution shape} of retrieval scores provide a reliable
enough signal without reading document content, or does
content-aware judging add essential information?

This question maps naturally to \emph{Query Performance Prediction}
(QPP)~\cite{arabzadeh2024tutorial}, a well-studied task in information
retrieval whose recent literature spans four threads relevant here:
classic score-statistic predictors, dense-retrieval extensions, direct
applications to RAG, and LLM-as-judge alternatives.

\paragraph{Classic post-retrieval QPP.}
QPP has a long history in sparse IR, with unsupervised
post-retrieval predictors derived from top-$k$ score statistics.
Three classics anchor the field: NQC~\cite{shtok2012nqc} reads
query drift from top-$k$ standard deviation; WIG~\cite{zhou2007wig}
contrasts top-$k$ mean against a corpus baseline;
SMV~\cite{tao2014smv} combines magnitude and dispersion. These
signals remain the backbone of later QPP studies, including those
below.

\paragraph{Neural and dense QPP.}
Dense retrievers break some classic-predictor assumptions.
Coherence-based measures~\cite{vlachou2024coherence} exploit
embedding-space neighbors; projection-displacement and
dimension-importance
estimators~\cite{faggioli2025pdqpp,faggioli2025dime,faggioli2023spatial}
probe dense-representation geometry. A cross-paradigm
evaluation~\cite{chifu2025limitations} finds classic predictors
fail to generalize across collections and retrievers, motivating
supervised feature-rich models.

\paragraph{QPP for RAG.}
Direct application of QPP to RAG is recent. Huly et~al.~\cite{huly2025sigir}
train a supervised regressor on six post-retrieval features (WIG, NQC,
SMV, top-$k$ mean and variance, U-SMV) to predict the perplexity gain
that retrieval provides for text completion. Tian
et~al.~\cite{tian2026ecir} define Retrieval and Generation Performance
Prediction, combining four score-only features (NQC, MaxScore, Dense-QPP,
A-Pair-Ratio) with BERT-QPP~\cite{arabzadeh2021bertqpp}, a cross-encoder
regressor on top-1 relevance. Both works confirm that retrieval-side features alone carry
substantial signal for RAG success, but neither compares them
against an LLM that actually reads the retrieved documents.

\paragraph{LLMs as retrieval judges.}
A complementary line uses LLMs to assess retrieval quality:
QPP-GenRE~\cite{meng2025qppgenre} fine-tunes open-source LLMs to
emit per-document relevance judgments reconstructable into
classical IR metrics, and Self-RAG~\cite{asai2024selfrag} uses
reflection tokens for self-critique. Such judges can be accurate
but inference cost scales with retrieval depth and context, and
sensitive deployments may disallow third-party APIs entirely.

\paragraph{Vision document retrieval.}
ColPali~\cite{faysse2024vidore} embeds document page images
directly via late-interaction MaxSim, outperforming OCR
pipelines. ViDoRe V3~\cite{mace2026vidorev3} benchmarks vision RAG
across eight enterprise domains. QPP for this multimodal
page-image setting remains relatively unexplored, particularly for
retrieval sufficiency in vision-document RAG.

\paragraph{Contributions.}
(i)~\emph{GeneralQPP}, a 24-feature non-lexical predictor for
retrieval sufficiency that reads no document content; (ii)~a
uniform-protocol comparison of score-based, content-based, and
hybrid paradigms on ViDoRe (8~vision domains) and a Czech
regulatory corpus (S\'{U}JB); (iii)~evidence that a local
multimodal LLM judge is more useful as an auxiliary feature than
as a standalone predictor, while score-distribution features
give the best latency--quality trade-off in our setting.

\section{Methods}
\label{sec:methods}

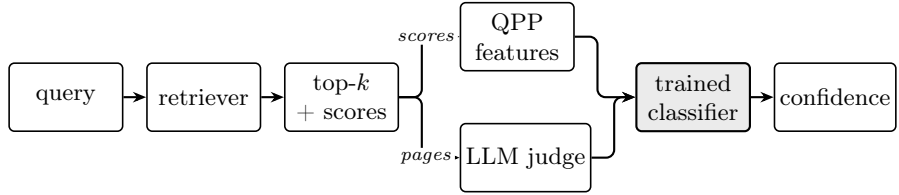
\begin{figure}[t]
\centering
\begin{tikzpicture}[
  font=\footnotesize,
  >={Stealth[length=5pt,width=4pt]},
  box/.style={draw, semithick, rounded corners=2pt, align=center,
              minimum width=15mm, minimum height=9mm, inner sep=2pt,
              fill=white},
  accent/.style={box, fill=black!8, thick},
  arr/.style={->, thick, rounded corners=3pt},
  lbl/.style={font=\scriptsize\itshape, inner sep=1.5pt, fill=white},
]
  \node[box]                                       (q)    {query};
  \node[box,    right=3mm of q]                    (ret)  {retriever};
  \node[box,    right=3mm of ret]                  (topk) {top-$k$\\+ scores};
  \node[box,    right=8mm of topk, yshift=8mm]     (qpp)  {QPP\\features};
  \node[box,    right=8mm of topk, yshift=-8mm]    (c1)   {LLM judge};
  \node[accent, right=8mm of qpp,  yshift=-8mm]    (clf)  {trained\\classifier};
  \node[box,    right=3mm of clf]                  (dec)  {confidence};

  \draw[arr] (q)   -- (ret);
  \draw[arr] (ret) -- (topk);
  \draw[arr] (topk.east) -- ++(3mm,0) |- (qpp.west)
                          node[lbl, pos=0.55]{scores};
  \draw[arr] (topk.east) -- ++(3mm,0) |- (c1.west)
                          node[lbl, pos=0.55]{pages};
  \draw[arr] (qpp.east)  -- ++(3mm,0) |- (clf.west);
  \draw[arr] (c1.east)   -- ++(3mm,0) |- (clf.west);
  \draw[arr] (clf)       -- (dec);
\end{tikzpicture}
\caption{Retrieval-sufficiency pipeline. Retriever:
\emph{Qwen3-VL-Embedding-8B} (4096-dim; text encoder on queries,
vision encoder on page images) returns the top-$k$ pages with
similarity scores. \textbf{Score-only methods (S0--S4)} read only
the score vector; the \textbf{content-based method C1}
(\emph{Qwen3.5-35B-A3B-FP8}, multimodal MoE: 35\,B total / 3\,B
active per token, FP8 weights, 262\,K-token context) reads the
retrieved JPEG pages (max~$1{,}568$\,px) and returns JSON with
reasoning, score~$\in[0,100]$, and a sufficient flag.
\textbf{Hybrids (H1--H3)} feed both signals into one trained
classifier whose confidence output drives the answer / abstain
decision in the downstream RAG generator.}
\label{fig:pipeline}
\end{figure}

We evaluate methods from three paradigms: score-based (S0--S4),
content-based (C1), and hybrid (H1--H3); the full pipeline is
shown in Fig.~\ref{fig:pipeline}. All trained methods
(S1--S4, H1--H3) share the architecture-selection protocol
described below (Section~\ref{sec:methods:arch}); per-method
feature sets are then specified in
Sections~\ref{sec:methods:score}--\ref{sec:methods:hybrid}.

\subsection{Architecture Selection}
\label{sec:methods:arch}

All trained methods (S1--S4, H1--H3) share a common protocol:
3-fold stratified CV on the training set selects the best architecture
by AUROC, features are standardized, and the chosen model is refit on
the full training fold. The base pool contains Logistic Regression,
Ridge~($\alpha\!=\!1.0$), and three MLPs up to~$(64,32)$ with early
stopping; for datasets with larger training sets, we extend with wider
MLPs~$(128,64)$ and~$(256,128,64)$ at $\alpha \in \{1.0, 0.1, 0.01\}$
(11~candidates total), fixed a~priori by training set size
(Section~\ref{sec:setup:data}). Per-method CV picks are reported in
Table~\ref{tab:wallclock}. C1 is not trained and has no architecture
to select; it uses only a monotonic Platt-scaling
wrapper~\cite{platt1999probabilistic} (Logistic Regression on the
scalar LLM score), which preserves AUROC and AUPRC and changes only
the calibration metrics (Brier, ECE). All preprocessing and
model selection are performed strictly within the training fold
of each split.

\subsection{Score-Based Methods}
\label{sec:methods:score}

\paragraph{S0: MaxSim Baseline.}
The simplest baseline uses the maximum similarity score as the confidence
estimate, with no learned model.

\paragraph{S1: GeneralQPP (Ours; \emph{GenQPP} in tables).}
We propose 24~non-lexical features organized in three groups
(Table~\ref{tab:s1_features}), trained on binary Hit@$k$ labels with
per-dataset architecture selection (Section~\ref{sec:methods:arch}).
The 24 features target distribution \emph{shape} (gaps, decay,
skewness, bimodality, concentration), query-surface statistics, and
global similarity-range baselines; only \emph{sim\_max} overlaps
with the classical pool (MaxScore). The key novelty is that most
features describe the geometry of the retrieval-score profile
itself rather than reusing classical IR score statistics verbatim.
We additionally report S1-Lean, a 13-feature subset selected by
greedy forward/backward stepwise search to maximize
Leave-One-Domain-Out (LODO) AUROC on ViDoRe
(Section~\ref{sec:discussion}): 8 distribution-shape features and
all 5 query-surface features (no global features survive selection).

\begin{table}[t]
\centering
\caption{GeneralQPP (S1) feature set: 24 non-lexical features in
three groups. Distribution features capture the \emph{shape} of
the similarity-score vector; query features are non-lexical text
statistics; global features summarize corpus-level similarity.
$\tau$ is a threshold calibrated for 90\% coverage; gap concentration
is (top1\,--\,top2)\,/\,(top1\,--\,top10), the share of the top-10
score range sitting in the first rank gap; bimodal gap is
$\mathrm{mean}(\text{top-3}) - \mathrm{mean}(\text{ranks 10\,--\,30})$;
top-5 concentration is $\sum\text{top-5}\,/\,\sum\text{top-100}$.
Features in \textbf{bold}
(8 distribution + 5 query) form the S1-Lean subset selected by LODO
forward-backward stepwise (Section~\ref{sec:discussion}).}
\label{tab:s1_features}
\setlength{\tabcolsep}{3pt}\footnotesize
\begin{tabular}{@{}lp{0.78\linewidth}@{}}
\toprule
Group & Features \\
\midrule
\makecell[lt]{Distribution\\(15)} &
  top1\,--\,p99 gap,
  \textbf{top1\,--\,top10 gap},
  \textbf{top-10 std.\ dev.},
  score decay slope,
  \textbf{bimodal gap},
  exp.\ decay rate,
  \textbf{99th percentile},
  $n$ above 0.8,
  \textbf{$n$ above 0.7},
  \textbf{top-50 skewness},
  \textbf{top-5 concentration},
  \textbf{max 2nd derivative (elbow)},
  $n$ above $\tau$,
  top-1 margin over $\tau$,
  gap concentration \\
\midrule
\makecell[lt]{Query\\(5)} &
  \textbf{char length},
  \textbf{word count},
  \textbf{has numbers},
  \textbf{punctuation count},
  \textbf{avg.\ word length} \\
\midrule
\makecell[lt]{Global\\(4)} &
  corpus mean sim.,
  corpus median sim.,
  max sim.,
  percentile range (p90\,--\,p10) \\
\bottomrule
\end{tabular}
\end{table}

\paragraph{S2: Huly et~al.~SIGIR'25~\cite{huly2025sigir}.}
Six features from the post-retrieval prediction framework: WIG, NQC, SMV,
top-$k$ mean, top-$k$ variance, and U-SMV.

\paragraph{S3: Tian et~al.~ECIR'26 (score-only)~\cite{tian2026ecir}.}
Four score-only features from the RPP framework are NQC, MaxScore, Dense-QPP,
and A-Pair-Ratio. We omit BERT-QPP (their fifth feature) as it reads
document content.

\paragraph{S4: Classic Full.}
A literature baseline pooling 18 score-only QPP features into one
trained predictor: Tian et~al.\ (NQC, MaxScore, Dense-QPP,
A-Pair-Ratio)~\cite{tian2026ecir}; Huly et~al.\ (WIG, SMV, U-SMV,
top-$k$ mean, top-$k$ variance)~\cite{huly2025sigir}; Faggioli
et~al.\ (query-centroid similarity, top-$k$ dispersion, query drift,
angular dispersion, $k$-NN distance, DIME-PRF concentration,
PRF-DIME)~\cite{faggioli2023spatial,faggioli2025dime,faggioli2025pdqpp};
Vlachou-Konchylaki et~al.\ (top-10 coherence, score-weighted
coherence)~\cite{vlachou2024coherence}. S4 controls for total
feature-set size when benchmarking against GeneralQPP.

\subsection{Content-Based Method}
\label{sec:methods:content}

\paragraph{C1: LLM Judge.}
C1 uses the multimodal MoE judge shown in
Fig.~\ref{fig:pipeline}~\cite{qwen2026qwen35}. Generation config:
\texttt{max\_tokens=512}, default temperature, thinking mode
disabled (enabling it caused parse failures). Confidence is
$\textsl{score}/100$; outputs that neither a multilingual JSON
parser (English and Czech keys) nor keyword/fraction fallbacks can
score default to~$0.5$. The same pipeline is used on ViDoRe and
S\'{U}JB; prompt, parser, and per-query raw scores are in the code
release.

\paragraph{Prompt selection.}
We use a three-shot English prompt (1~positive, 2~negative
calibration examples), selected because it achieved zero parse
failures and the best overall AUROC among four tested variants
(Table~\ref{tab:c1_prompt_selection}); a Czech domain-specific
variant gave no ranking benefit.

\begin{table}[t]
\centering
\caption{Vision prompt variants for C1 (stratified $n=200$
S\'{U}JB train, Hit@10). \emph{Overall}: unparseable outputs
score~$0.5$; \emph{Parseable}: excludes them. $p$\,vs.\,chosen:
paired bootstrap AUROC test against the few-shot chosen variant
(2{,}000~resamples).}
\label{tab:c1_prompt_selection}
\setlength{\tabcolsep}{5pt}\scriptsize
\begin{tabular}{@{}lcccc@{}}
\toprule
Variant & Overall & Parseable & Fails & $p$\,vs.\,chosen \\
\midrule
zero-shot         & 0.866 & 0.885 & 32 & 0.39 \\
CoT               & 0.820 & 0.867 & 46 & $<$0.001 \\
few-shot (chosen) & 0.884 & 0.884 &  0 & -- \\
few-shot S\'{U}JB & 0.886 & 0.886 &  0 & 0.94 \\
\bottomrule
\end{tabular}
\end{table}

\subsection{Hybrid Methods}
\label{sec:methods:hybrid}

\paragraph{H1: Tian et~al.~ECIR'26 (adapted)~\cite{tian2026ecir}.}
Tian's full method augments S3's four score-only features with a
cross-encoder relevance score for the top-1 retrieved document.
The original uses BERT-QPP~\cite{arabzadeh2021bertqpp}, a BERT
cross-encoder fine-tuned on relevance judgments to regress MRR. We
replace it with a \emph{zero-shot} multilingual cross-encoder
(\texttt{mmarco-mMiniLMv2-L12-H384-v1}) because our S\'{U}JB dataset
(Section~\ref{sec:setup:data}) contains Czech queries and no Czech
QPP-labeled training data is available, so an English-fine-tuned
BERT-QPP would not transfer. H1 should therefore be read as a
\emph{conservative} approximation of Tian et~al.'s full hybrid
(a fully fine-tuned BERT-QPP would strengthen it). Total: 5~features.

\paragraph{H2: GeneralQPP + LLM Judge (Ours).}
Our 24~GeneralQPP features augmented with the C1 sufficiency score as a
25th feature. This treats the LLM judgment as an additional signal for the
trained predictor rather than a standalone estimator.

\paragraph{H3: Classic Full + LLM Judge.}
The 18~classic QPP features (S4) augmented with the C1 score as a 19th
feature.

\begin{table}[t]
\centering
\caption{Per-query wallclock latency and CV-selected
architectures. S0--S4/H1 measured on the S\'{U}JB test split
(302~queries, 5~repetitions, mean); C1 end-to-end over the
full 1510-query run at $k\!=\!10$; H2/H3 are additive (S-method
+ one C1 call). Arch: R=Ridge~($\alpha\!=\!1.0$), L=Logistic,
MLP=$(256,128,64)$; ViDoRe entries list per-domain picks over
8~domains.}
\label{tab:wallclock}
\setlength{\tabcolsep}{3pt}\scriptsize
\begin{tabular}{@{}llcllc@{}}
\toprule
Tier & Method & Ft & Arch\,(SÚJB) & Arch\,(ViDoRe) & Latency/query \\
\midrule
S0 & MaxSim baseline            & 1  & --    & --                                    & 7\,$\mu$s \\
S1 & GenQPP (ours)              & 24 & R     & 8$\times$MLP                          & 2.0\,ms \\
S1-Lean & GenQPP-Lean (ours)     & 13 & L     & 8$\times$MLP                          & 1.9\,ms \\
S2 & Huly SIGIR'25              & 6  & L     & 7$\times$MLP, 1$\times$R              & 2.8\,ms \\
S3 & Tian ECIR'26 (score-only)  & 4  & L     & 5$\times$MLP, 2$\times$R, 1$\times$L  & 2.7\,ms \\
S4 & Classic Full               & 18 & R     & 8$\times$MLP                          & 2.7\,ms \\
\midrule
C1 & LLM judge (Qwen3.5-35B)    & 1  & Platt & Platt                                 & 6.3\,s \\
\midrule
H1 & Tian ECIR'26 (adapted)     & 5  & L     & 6$\times$MLP, 1$\times$R, 1$\times$L  & 982.9\,ms \\
H2 & GenQPP + LLM (ours)        & 25 & L     & 8$\times$MLP                          & 6.3\,s \\
H3 & Classic Full + LLM (ours)  & 19 & R     & 8$\times$MLP                          & 6.3\,s \\
\bottomrule
\end{tabular}
\end{table}

\section{Experimental Setup}
\label{sec:setup}

\subsection{Datasets}
\label{sec:setup:data}

We evaluate on two
datasets\footnote{Code and data: \url{https://github.com/veselm73/rag-confidence}.}:
S\'{U}JB (in-domain, Czech regulatory) and
ViDoRe (cross-domain, 8~domains).

\paragraph{S\'{U}JB.}
The production RAG corpus motivating this work: 517~pages from
5~Czech nuclear regulatory documents with 1{,}510 synthetic queries
generated by GPT-4o from page-level OCR (stratified per document):
510 standard answerable grounded in one page, 500 paraphrases
(synonym, full-rephrase, cross-lingual), and 500 adversarial
unanswerable (Section~\ref{sec:setup:gt} defines the negative
classes). We use synthetic queries because real user logs from the
deployment were not available during development; we therefore
treat S\'{U}JB primarily as a controlled deployment case study
rather than a substitute for live-query evaluation. We use binary
sufficient/not-sufficient labels: a page is \emph{sufficient} iff
the query is fully answerable from its content. Retriever and
judge are as in Fig.~\ref{fig:pipeline}. We use a fixed 80/20 split
(1{,}208 train, 302 test, seed=42) with the 5-candidate architecture
pool of Section~\ref{sec:methods:arch}.

\paragraph{ViDoRe.}
For ViDoRe V3~\cite{mace2026vidorev3} we use an 8-domain benchmark
setup (physics, energy, finance~EN/FR, HR, industrial, computer
science, pharmaceuticals) comprising 14{,}514~queries; retriever
as in Fig.~\ref{fig:pipeline}. V3 uses human-verified relevance
labels 1~(relevant) or 2~(fully relevant, self-sufficient);
unlabeled pages are treated as 0 (Section~\ref{sec:setup:gt}
defines the negative classes). We use stratified 80/20 splits within
each domain (seed=42), training and evaluating \emph{per} domain
to measure in-domain QPP performance across diverse retrieval
settings. Per-domain training
sets range from 1{,}032 to 1{,}747 queries, supporting the extended
11-candidate pool of Section~\ref{sec:methods:arch}.

\subsection{Ground Truth}
\label{sec:setup:gt}

Retrieval success is Hit@$k$: positive if at least one
self-sufficient page appears in the top-$k$ (\emph{score~$=1$} on
S\'{U}JB, \emph{score~$=2$} on ViDoRe). Two kinds of
negatives: ViDoRe \emph{unanswerable} (no score-2 page in the
corpus; score-1 partial matches may still exist) and S\'{U}JB
\emph{adversarial} (topic absent from the corpus entirely, a
stricter failure mode). We report both $k=5$ and $k=10$ as a
cutoff-robustness check; ViDoRe's eight domains test
cross-distribution generalization
(Section~\ref{sec:results:consistency}). Hit@$k$ matches the
operational RAG objective in our setting: the system needs at
least one self-sufficient page to support a grounded answer, not
majority recall~\cite{huly2025sigir}.

\subsection{Evaluation Metrics}

Each method produces a confidence score $\hat{p}_i \in [0,1]$ for
query~$q_i$. We evaluate against binary ground-truth labels
$y_i \in \{0,1\}$ using four complementary metrics: AUROC
(discrimination: ability to rank successes above failures), AUPRC
(precision--recall, sensitive to class imbalance), Brier
score~\cite{brier1950verification} (calibration quality, lower is
better), and ECE~\cite{naeini2015ece} (expected calibration error
over 10~equal-width bins, lower is better). AUROC is our primary
selection and comparison metric; AUPRC, Brier, and ECE serve as
complementary diagnostics. All results are averaged over 5~random
seeds for model initialization.

\section{Results}
\label{sec:results}

\subsection{Main Results: ViDoRe Benchmark}
\label{sec:results:main}

On the eight ViDoRe domains (14{,}514~queries total) we train one
model per domain on the 80\% in-domain split, with architecture
selected by 3-fold CV from the 11-candidate pool of
Section~\ref{sec:methods:arch}, and evaluate on the 20\% held-out
test. Table~\ref{tab:vidore_per_domain} reports per-domain AUROC
and the weighted average, our primary headline result.

\begin{table}[t]
\centering
\caption{Per-domain test AUROC on the eight ViDoRe in-domain
settings (rounded to three decimals); the right-most column is the
weighted average across all eight domains. Stratified paired-bootstrap 95\% CIs on the weighted
average and headline pairwise $p$-values are reported in the
accompanying \S\ref{sec:results:main} text (2{,}000~resamples,
cluster-resampled within each domain). \textbf{Bold}: best
score-only method per domain; \underline{underlined}: best overall
per domain. S1-Lean is the LODO-robust subset
(Section~\ref{sec:discussion}), included here for completeness; its
intended regime is cross-domain transfer.}
\label{tab:vidore_per_domain}
\setlength{\tabcolsep}{3pt}\footnotesize
\begin{tabular}{@{}lcccccccc|c@{}}
\toprule
 & \rotatebox{60}{cs} & \rotatebox{60}{energy} & \rotatebox{60}{fin\,en} & \rotatebox{60}{fin\,fr} & \rotatebox{60}{hr} & \rotatebox{60}{indust} & \rotatebox{60}{pharma} & \rotatebox{60}{physics} & Avg \\
$n_\textit{test}$ & 258 & 370 & 371 & 384 & 382 & 340 & 437 & 363 & 2{,}905 \\
\midrule
S0 & .589 & .573 & .636 & .644 & .536 & .497 & .615 & .639 & .593 \\
S3 & .585 & .583 & .709 & .610 & .706 & .649 & .634 & .655 & .643 \\
S2 & .756 & .520 & .720 & .780 & .662 & .756 & .801 & .716 & .714 \\
S4 & .764 & \textbf{.835} & .822 & .849 & .838 & .840 & \underline{\textbf{.869}} & .837 & .835 \\
S1 (ours) & \underline{\textbf{.905}} & .787 & \textbf{.826} & \underline{\textbf{.889}} & \textbf{.848} & .858 & .854 & \underline{\textbf{.900}} & \textbf{.856} \\
S1-Lean (13 ft, ours) & .809 & .748 & .797 & .776 & .824 & \underline{\textbf{.862}} & .731 & .730 & .782 \\
\midrule
C1 & .627 & .654 & .724 & .687 & .608 & .693 & .664 & .530 & .649 \\
\midrule
H1 & .641 & .584 & .668 & .714 & .697 & .697 & .748 & .686 & .683 \\
H3 & .887 & .845 & .826 & .762 & \underline{.864} & .835 & .862 & .816 & .836 \\
H2 (ours) & .880 & \underline{.906} & \underline{.846} & .857 & .834 & .856 & .853 & .880 & \underline{.863} \\
\bottomrule
\end{tabular}
\end{table}

\paragraph{Score-based methods.}
S1 (GeneralQPP) achieves the best weighted-average AUROC of 0.856
(95\% CI [0.842, 0.870]) among score-only methods, ahead of S4
($+0.021$, S1 wins 6/8 domains), S2 ($+0.142$), and S3 ($+0.213$);
all three significant under a stratified paired bootstrap
(2{,}000~resamples, cluster-resampled per domain; S1$>$S4 $p=0.012$,
S1$>$S2/S3 $p=0.001$). This makes S1 the strongest standalone
score-only predictor on ViDoRe, while H2 provides the best absolute
AUROC at LLM-level latency (below). S4 wins on energy and pharma
(by 0.015--0.048), a reasonable fallback when training data is limited.

\paragraph{Content-based method.}
C1 (LLM judge) achieves AUROC~0.649 [0.628, 0.670], below all
trained score-only methods except S0. The $+0.207$ gap to S1 is
large and statistically significant (paired bootstrap
$[+0.184, +0.231]$, $p=0.001$); C1 also pays $\sim$3000$\times$ more
per query (Table~\ref{tab:wallclock}).

\paragraph{Hybrid methods.}
H2 (GeneralQPP + LLM judge) achieves AUROC 0.863 [0.849, 0.876],
$+0.007$ over S1 (not significant on ViDoRe: bootstrap
$[-0.005, +0.018]$, $p=0.261$), concentrated in the energy domain
(H2~0.906 vs.\ S1~0.787), where the LLM judge compensates for the
weakest score-feature domain; on S\'{U}JB
(Section~\ref{sec:results:sujb}), H2$>$S1 reaches $\alpha=0.05$
at both retrieval depths ($p=0.031$ at Hit@5, $p=0.023$ at
Hit@10). H2 beats H3 (Classic + LLM) by
$+0.028$ [$+0.012, +0.042$], $p=0.002$, showing that GeneralQPP
contributes a signal that the classic pool does not provide, even
after the LLM score. H1 (Tian adapted, zero-shot cross-encoder)
reaches only 0.683 [0.662, 0.702]; fine-tuning would likely improve
this. In practical terms, ViDoRe therefore favors S1 as the
latency-efficient choice, and H2 only when an LLM pass is already
acceptable.

\subsection{Diagnostic: Failure-Case Calibration}
\label{sec:results:adv_detection}

\begin{figure}[t]
  \centering
  \includegraphics[width=\linewidth]{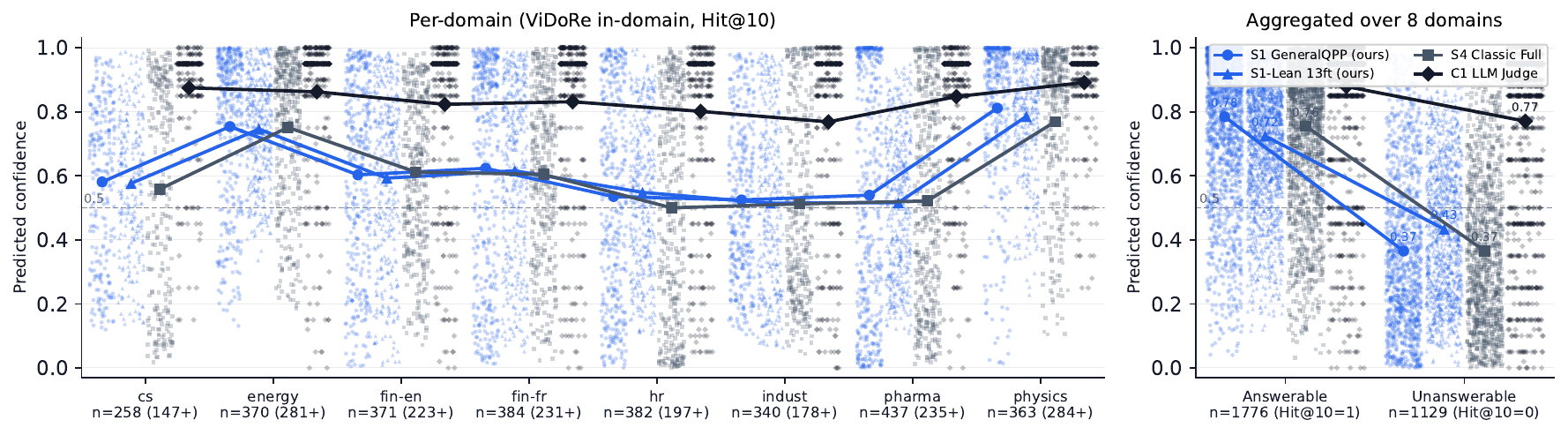}
  \caption{Per-query predicted confidence on the eight ViDoRe
  in-domain 80/20 test splits (Hit@10, mean of five seeds).
  \emph{Left:} per-domain; each dot is one test query, solid
  lines connect method means. \emph{Right:} same queries
  aggregated across all eight domains and split by ground-truth
  Hit@10; annotations are per-group means (2~decimals), whose
  positive$-$negative differences appear in
  \S\ref{sec:results:adv_detection}. The key pattern is that S1
  separates answerable from unanswerable queries far more cleanly
  than C1, especially in the aggregated panel.}
  \label{fig:stripplot_vidore}
\end{figure}

On ViDoRe, queries with Hit@10=0 range from 21.8\%
(physics) to 48.4\% (HR). C1 assigns mean confidence~0.560 to
these queries after the Platt wrapper
(Section~\ref{sec:methods:arch}) and~0.771 raw, both exceeding
MaxSim's~0.467 and confirming systematic LLM overconfidence when
the retriever fails. Score-based methods assign substantially
lower confidence (S1 and S4 both~0.365); H2 reaches~0.355, the
lowest of all methods, showing that the trained predictor
downweights C1's overoptimistic scores. Weighted-average
positive-minus-negative confidence separation reinforces the
pattern: C1~(0.047) vs.\ S1~(0.394) vs.\ H2~(0.418). Platt
calibration \emph{shrinks} C1's separation from 0.098 raw
to 0.047, making its weak discrimination visible
(Fig.~\ref{fig:stripplot_vidore}). For deployment, this matters
more than headline AUROC alone: overconfident failure cases are
precisely the queries on which abstention is needed.

\subsection{Deployment Case Study: S\'{U}JB}
\label{sec:results:sujb}

Table~\ref{tab:sujb} reports full S\'{U}JB results (302~test
queries, 53.0\% positive rate at Hit@5) at both retrieval depths.
This deployment-oriented case study tests how the methods behave
on 500~synthetic adversarial unanswerable queries under
a realistic local-regulatory setup.

\begin{table}[t]
\centering
\caption{S\'{U}JB results in vision mode (302~test queries, mean over
5~seeds) at both Hit@5 and Hit@10 retrieval depths. Best score-only in
\textbf{bold}, best overall \underline{underlined}. C1 values are
Platt-calibrated (Section~\ref{sec:results:c1p}).}
\label{tab:sujb}
\setlength{\tabcolsep}{2pt}\scriptsize
\begin{tabular}{@{}llc|cccc|cccc@{}}
\toprule
& & & \multicolumn{4}{c|}{Hit@5} & \multicolumn{4}{c}{Hit@10} \\
\cmidrule(lr){4-7} \cmidrule(l){8-11}
Tier & Method & Ft & AUROC & AUPRC & Brier & ECE & AUROC & AUPRC & Brier & ECE \\
\midrule
\multirow{6}{*}{\rotatebox{90}{\scriptsize Score}}
& S0 MaxSim              & 1  & .552 & .575 & .262 & .118 & .545 & .612 & .270 & .164 \\
& S3 Tian ECIR'26        & 4  & .815 & .813 & .183 & .101 & .807 & .831 & .185 & .100 \\
& S2 Huly SIGIR'25       & 6  & .871 & .877 & .147 & .059 & .875 & .905 & .144 & .067 \\
& S4 Classic Full        & 18 & .875 & .868 & .145 & .059 & .891 & .908 & .131 & .048 \\
& S1 GenQPP (ours)       & 24 & \textbf{.893} & .893 & .136 & .055 & \textbf{.905} & .923 & .124 & .061 \\
& S1-Lean (ours)         & 13 & .890 & \textbf{.894} & \textbf{.135} & \underline{\textbf{.033}} & .904 & \textbf{.923} & \textbf{.124} & \underline{\textbf{.043}} \\
\midrule
Cont.
& C1 LLM Judge           & 1  & .832 & .831 & .168 & .061 & .876 & .898 & .140 & .074 \\
\midrule
\multirow{3}{*}{\rotatebox{90}{\scriptsize Hybrid}}
& H1 Tian (adapted)      & 5  & .820 & .820 & .179 & .087 & .809 & .839 & .184 & .102 \\
& H3 Classic + LLM       & 19 & .906 & .895 & .125 & .074 & \underline{.932} & .933 & \underline{.096} & .075 \\
& H2 GenQPP + LLM (ours) & 25 & \underline{.911} & \underline{.904} & \underline{.120} & .071 & .931 & \underline{.935} & .101 & .076 \\
\bottomrule
\end{tabular}
\end{table}

\paragraph{ViDoRe ranking replicates on S\'{U}JB.}
The ranking observed on ViDoRe broadly replicates on S\'{U}JB: S1
leads the score-only tier at AUROC~0.893 and improves over S2,
S4, and C1 on all four reported metrics. Because the S\'{U}JB test
set contains only 302~queries, we treat it primarily as a
deployment case study and interpret statistical significance
cautiously; the main inferential evidence comes from the much
larger ViDoRe benchmark. In sum: on S\'{U}JB, S1 wins
score-only; H2 wins on Hit@5 across three of four metrics, while H3
is competitive at Hit@10; S1-Lean provides the best Hit@10
calibration.

\paragraph{Performance by query type.}
The 302~test queries split into 102~standard answerable,
100~paraphrased (synonym, full-rephrase, and cross-lingual
variants), and 100~adversarial unanswerable. This breakdown
matters operationally because adversarial unanswerables are the
highest-risk queries for a production assistant.
Table~\ref{tab:query_types} reports AUROC on the 202~answerable
queries and adversarial-detection AUROC (answerable vs.\
unanswerable).

\begin{table}[t]
\centering
\caption{S\'{U}JB per-query-type breakdown (302~test queries,
mean over 5~seeds). \emph{Ans.\ AUROC}: among the 202~answerable
queries only. \emph{Adv.\ AUROC}: separation of answerable vs.\
unanswerable. $\bar{p}_\text{ans}/\bar{p}_\text{adv}$: mean
predicted confidence on answerable/adversarial queries. Best
score-only \textbf{bold}, best overall \underline{underlined}.}
\label{tab:query_types}
\setlength{\tabcolsep}{2pt}\scriptsize
\begin{tabular}{@{}ll|cccc|cccc@{}}
\toprule
& & \multicolumn{4}{c|}{Hit@5} & \multicolumn{4}{c}{Hit@10} \\
\cmidrule(lr){3-6} \cmidrule(l){7-10}
Tier & Method
  & \makecell{Ans.\\AUROC} & \makecell{Adv.\\AUROC} & $\bar{p}_\text{ans}$ & $\bar{p}_\text{adv}$
  & \makecell{Ans.\\AUROC} & \makecell{Adv.\\AUROC} & $\bar{p}_\text{ans}$ & $\bar{p}_\text{adv}$ \\
\midrule
\multirow{6}{*}{\rotatebox{90}{\small Score}}
& S0 MaxSim            & .636 & .489 & .410 & .417 & .665 & .489 & .410 & .417 \\
& S3 Tian ECIR'26      & .753 & .797 & .563 & .343 & .727 & .806 & .590 & .335 \\
& S2 Huly SIGIR'25     & \textbf{.771} & .887 & .648 & .245 & .715 & .908 & .691 & .237 \\
& S4 Classic Full      & .770 & .890 & .660 & .259 & .734 & .922 & .708 & .213 \\
& S1 GenQPP (ours)     & .764 & \textbf{.934} & .670 & .210 & .734 & \textbf{.949} & \textbf{.738} & .240 \\
& S1-Lean (ours)       & .764 & .931 & \textbf{.677} & \textbf{.166} & \textbf{.741} & .943 & .724 & \textbf{.164} \\
\midrule
Cont.
& C1 LLM Judge         & .683 & .864 & \underline{.748} & .270 & .742 & .886 & \underline{.801} & .282 \\
\midrule
\multirow{3}{*}{\rotatebox{90}{\small Hyb.}}
& H1 Tian (adapted)         & .769 & .788 & .563 & .337 & .735 & .804 & .590 & .332 \\
& H3 Classic + LLM          & .771 & .939 & .687 & .177 & \underline{.786} & .960 & .745 & \underline{.128} \\
& H2 GenQPP + LLM (ours)    & .770 & \underline{.954} & .692 & \underline{.154} & .759 & \underline{.970} & .764 & .181 \\
\bottomrule
\end{tabular}
\end{table}

\begin{figure}[t]
  \centering
  \includegraphics[width=\linewidth]{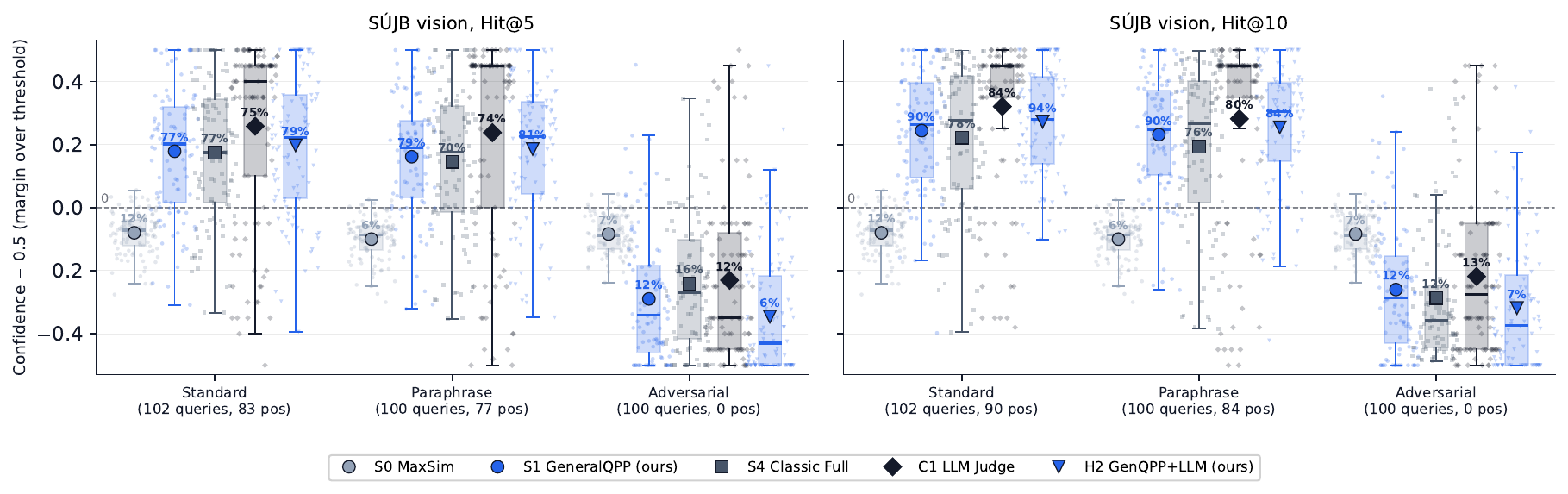}
  \caption{Per-query predicted confidence on the S\'{U}JB
  vision-mode test set across query types at Hit@5
  (\emph{left}) and Hit@10 (\emph{right}). S1 and S4
  (score-only) drop below the 0.5 threshold on adversarial
  queries; C1 stays over-confident; S0 is flat across all three
  groups. The main operational takeaway is that score-based
  predictors fall below the 0.5 threshold on adversarial queries
  much more reliably than the standalone LLM judge.}
  \label{fig:stripplot_sujb}
\end{figure}

S1 beats C1 on adversarial detection (0.934 vs.\ 0.864) despite
seeing only similarity scores (Fig.~\ref{fig:stripplot_sujb}):
unanswerable queries produce characteristically flat score
distributions that shape features capture reliably, whereas S0 is
near-random (0.489). Among answerable queries all trained methods
cluster (0.753--0.771), so the LLM judge's value lies mostly in
in-scope discrimination.

\paragraph{Operating-threshold view.}
The \% annotations in Fig.~\ref{fig:stripplot_sujb} give each
method's would-answer rate at $\tau=0.5$. On the 100~adversarial
queries, H2 crosses only 6--7\% at Hit@5/10 vs.\ 12\% for S1/C1
and 12--16\% for S4, roughly halving the best classic baseline's
false-answer rate. On answerables, H2's rates (79--94\%) track ground-truth
positives (77--88\%): H2 abstains where the pipeline should reject
and answers elsewhere, the asymmetry safety-critical RAG requires.

\subsection{Robustness Checks}
\label{sec:results:consistency}\label{sec:results:c1p}\label{sec:results:overfit}

Method rankings agree across datasets (Spearman
$\rho = 0.90$, Kendall $\tau = 0.78$, $p < 0.005$ between the
ViDoRe weighted-average and S\'{U}JB Hit@10 AUROC columns),
suggesting that the broad method ranking is stable across the
benchmark and deployment case study despite differences in
language, document type, and annotation. Platt calibration cuts
C1's ECE $6\times$ on ViDoRe (0.249$\to$0.041) and halves it on
S\'{U}JB Hit@5 (0.118$\to$0.061) without changing AUROC, so the
$+0.207$ S1$-$C1 gap reflects discrimination, not calibration.
Train--test AUROC gaps stay below~0.023 for every method reported
in Tables~\ref{tab:vidore_per_domain}--\ref{tab:sujb}, well below
common practical overfitting concern thresholds.

\section{Discussion}
\label{sec:discussion}

\paragraph{Score Features vs.\ LLM Judges.}
S1 beats C1 by $+0.207$ weighted-average AUROC on ViDoRe (winning
all eight domains) and by $+0.061$ on S\'{U}JB Hit@5, despite
reading no document content at 2\,ms per query compared to C1's
$\sim$6.3\,s LLM pass. Section~\ref{sec:results:adv_detection}
further shows C1 stays over-confident on ViDoRe failures while
score-distribution methods back off appropriately. On ViDoRe, S1
offers the stronger latency--quality trade-off, as H2's AUROC
gain over S1 is small and statistically non-significant while
requiring LLM-level latency.

\paragraph{LLM as Feature, Not Predictor.}
On S\'{U}JB, H2 improves on both S1 and C1 in AUROC at both
retrieval depths (Table~\ref{tab:sujb}), while H3 is essentially
tied and slightly higher at Hit@10 AUROC; H2 also beats the
standalone LLM judge by a wide and significant margin on ViDoRe
($+0.214$, $p=0.001$).
The hybrid uplift over S1 is small and setting-dependent
($+0.018$ on S\'{U}JB Hit@5 vs.\ $+0.007$ on ViDoRe, $p=0.261$),
so H2 pays off primarily when the judge is already in the
inference path. To our knowledge, this is among the first
direct empirical comparisons in retrieval-sufficiency QPP for
vision RAG to show that, for our tested local multimodal judge,
the LLM is more effective as an auxiliary feature than as a
standalone predictor.

\paragraph{OOD Robustness: S1-Lean.}
LODO on ViDoRe drops S1 from 0.856 to 0.706 (0.15 penalty).
Forward-backward stepwise peaks at 13 features (LODO AUROC 0.719,
\emph{+0.012 over full S1 and +0.032 over classical baseline S4});
S1-Lean trades 0.074 in-domain ViDoRe AUROC for this transfer and
for the best SÚJB calibration of any method (ECE 0.043 at Hit@10
vs.\ S4's 0.048, H2's 0.076). A practical default is to use S1
when modest target-domain labels exist, and S1-Lean when
cross-domain transfer matters more.

\paragraph{Limitations.}
GeneralQPP features were designed on SÚJB training data; ViDoRe
mitigates this (S1 wins 6/8 domains) and the S1-Lean LODO result
quantifies cross-domain transfer. Validation on live
(non-synthetic) queries is the main item of future work
(Section~\ref{sec:setup:data}). With a single retriever and a
single local judge in our setup, retriever-agnosticism of the
distribution-shape features is an empirical prediction we have
not tested directly. C1 is a single local multimodal judge
constrained by JSON parsing, and our H1 is a conservative
adaptation (zero-shot cross-encoder in place of a fine-tuned
BERT-QPP); a stronger judge or fine-tuned hybrid could narrow
the gaps. Accordingly, our conclusions should be read
as strongest for local multimodal retrieval-sufficiency
prediction with modest target-domain supervision, not as a
universal ranking of all RAG confidence estimators.

\section{Conclusion}
\label{sec:conclusion}

We compared score-based, content-based, and hybrid QPP for RAG
retrieval sufficiency under a uniform protocol across nine methods
and two datasets (ViDoRe, 14{,}514 queries; S\'{U}JB,
1{,}510~queries). Our \textbf{GeneralQPP} reaches a weighted-average
AUROC of~0.856 on ViDoRe ($+0.207$ over a local Qwen3.5 judge) at
2\,ms per query, and an adversarial-detection AUROC of~0.934 on
S\'{U}JB. A \textbf{hybrid LLM-as-feature} method (H2) wins the
most evaluated settings (0.863 on ViDoRe, n.s.\ vs.\ S1;
0.911/0.954 on S\'{U}JB), with H3 essentially tied and slightly
higher on S\'{U}JB Hit@10 AUROC, showing that our local multimodal
judge contributes more as a feature than as a standalone predictor. Rankings agree across datasets
(Spearman~$\rho = 0.90$); to the best of our knowledge, this is
one of the first applications of QPP-style retrieval-sufficiency
prediction to vision-document retrieval. These conclusions are
strongest for local multimodal retrieval-sufficiency prediction
with modest target-domain supervision, under one tested
retriever/judge pairing; stronger judges and broader
answer-quality predictors may alter the ranking. Future work:
LLM-reasoning-trace attention, cross-lingual evaluation,
live-query validation, cross-retriever generalization, and
selective RAG with abstention.


\bibliographystyle{splncs04}
\bibliography{refs}

@inproceedings{lewis2020rag,
  author    = {Lewis, P. and others},
  title     = {Retrieval-augmented generation for knowledge-intensive {NLP} tasks},
  booktitle = {NeurIPS},
  year      = {2020}
}

@misc{arabzadeh2024tutorial,
  author       = {Arabzadeh, N. and Meng, C. and Aliannejadi, M. and Bagheri, E.},
  title        = {Query performance prediction: From fundamentals to advanced techniques},
  howpublished = {ECIR Tutorial},
  year         = {2024}
}

@inproceedings{tian2026ecir,
  author    = {Tian, S. and Ganguly, D. and Macdonald, C.},
  title     = {Predicting retrieval utility and answer quality in {RAG}},
  booktitle = {ECIR},
  year      = {2026}
}

@inproceedings{huly2025sigir,
  author    = {Huly, O. and Carmel, D. and Kurland, O.},
  title     = {Predicting {RAG} performance for text completion},
  booktitle = {SIGIR},
  year      = {2025}
}

@article{shtok2012nqc,
  author  = {Shtok, A. and Kurland, O. and Carmel, D. and Raiber, F. and Markovits, G.},
  title   = {Predicting query performance by query-drift estimation},
  journal = {ACM TOIS},
  volume  = {30},
  number  = {2},
  year    = {2012}
}

@inproceedings{zhou2007wig,
  author    = {Zhou, Y. and Croft, W. B.},
  title     = {Query performance prediction in web search environments},
  booktitle = {SIGIR},
  year      = {2007}
}

@inproceedings{tao2014smv,
  author    = {Tao, Y. and Wu, S.},
  title     = {Query performance prediction by considering score magnitude and variance together},
  booktitle = {CIKM},
  year      = {2014}
}

@inproceedings{faggioli2023spatial,
  author    = {Faggioli, G. and Ferro, N. and Muntean, C. I. and Perego, R. and Tonellotto, N.},
  title     = {Query performance prediction for neural {IR}: Are we there yet?},
  booktitle = {SIGIR},
  year      = {2023}
}

@inproceedings{vlachou2024coherence,
  author    = {Vlachou-Konchylaki, M. and Macdonald, C.},
  title     = {Coherence-based query performance measures for dense retrieval},
  booktitle = {ICTIR},
  year      = {2024}
}

@inproceedings{faggioli2025dime,
  author    = {Faggioli, G. and Ferro, N. and Perego, R. and Tonellotto, N.},
  title     = {Query performance prediction using dimension importance estimators},
  booktitle = {ECIR},
  year      = {2025}
}

@article{faggioli2025pdqpp,
  author  = {Faggioli, G. and Ferro, N. and Perego, R. and Tonellotto, N.},
  title   = {Projection-displacement-based {QPP} for dense retrievers},
  journal = {ACM TOIS},
  year    = {2025}
}

@article{chifu2025limitations,
  author  = {Chifu, A. G. and Dejean, S. and Mothe, J. and Garouani, M. and Ortiz, D. and Ullah, M. Z.},
  title   = {Uncovering the limitations of query performance prediction},
  journal = {ACM TOIS},
  year    = {2025}
}

@article{meng2025qppgenre,
  author  = {Meng, C. and Arabzadeh, N. and Askari, A. and Aliannejadi, M. and de Rijke, M.},
  title   = {{QPP} using relevance judgments generated by {LLMs}},
  journal = {ACM TOIS},
  year    = {2025}
}

@inproceedings{asai2024selfrag,
  author    = {Asai, A. and Wu, Z. and Wang, Y. and Sil, A. and Hajishirzi, H.},
  title     = {{Self-RAG}: Learning to retrieve, generate, and critique through self-reflection},
  booktitle = {ICLR},
  year      = {2024}
}

@inproceedings{arabzadeh2021bertqpp,
  author    = {Arabzadeh, N. and Khodabakhsh, M. and Bagheri, E.},
  title     = {{BERT-QPP}: Contextualized pre-trained transformers for {QPP}},
  booktitle = {CIKM},
  year      = {2021}
}

@misc{qwen2026qwen35,
  author       = {{Qwen Team}},
  title        = {{Qwen3.5}: Towards native multimodal agents},
  howpublished = {Qwen blog},
  year         = {2026},
  month        = feb,
  note         = {\url{https://qwen.ai/blog?id=qwen3.5}}
}

@misc{mace2026vidorev3,
  author       = {Mace, T. and others},
  title        = {{ViDoRe V3}: Evaluation of {RAG} in real-world scenarios},
  howpublished = {arXiv:2601.08620},
  year         = {2026}
}

@inproceedings{faysse2024vidore,
  author    = {Faysse, M. and others},
  title     = {{ColPali}: Efficient document retrieval with vision language models},
  booktitle = {ICLR},
  year      = {2025}
}

@article{brier1950verification,
  author  = {Brier, G. W.},
  title   = {Verification of forecasts expressed in terms of probability},
  journal = {Monthly Weather Review},
  volume  = {78},
  number  = {1},
  pages   = {1--3},
  year    = {1950}
}

@inproceedings{naeini2015ece,
  author    = {Naeini, M. P. and Cooper, G. F. and Hauskrecht, M.},
  title     = {Obtaining well calibrated probabilities using {Bayesian} binning},
  booktitle = {AAAI},
  year      = {2015}
}

@incollection{platt1999probabilistic,
  author    = {Platt, J. C.},
  title     = {Probabilistic outputs for support vector machines},
  booktitle = {Advances in Large Margin Classifiers},
  volume    = {10},
  number    = {3},
  pages     = {61--74},
  year      = {1999}
}

\end{document}